\documentstyle[11pt]{article}
\input epsf
\begin{document}
\thispagestyle{empty}
\vfill
\centerline{\LARGE Discreteness of Area and  Volume}
\centerline{\LARGE in Quantum Gravity}
\rm
\vskip1cm
\centerline{Carlo Rovelli${}^*$ and Lee Smolin${}^\dagger$}
\vskip .3cm
\centerline{\it ${}^*$ Department of Physics, University of
        Pittsburgh, Pittsburgh, Pa15260}
\vskip .3cm\vskip .3cm
 \centerline{\it ${}^\dagger$  Center for Gravitational Physics and
Geometry, Department of Physics}
\centerline{\it Pennsylvania State University,
University Park, Pa16802-6360}
 \vfill
\centerline{\today}
\vfill

\centerline{\bf Abstract}

\noindent
We study the operator that corresponds to the measurement of
volume, in non-perturbative quantum gravity, and we compute its
spectrum.  The operator is constructed in the loop representation, via
a regularization procedure; it is finite, background independent, and
diffeomorphism-invariant, and therefore well defined on the space of
diffeomorphism invariant states (knot states).  We find that the
spectrum of the volume of any physical region is discrete.  A family
of eigenstates are in one to one correspondence with the spin
networks, which were introduced by Penrose in a different context.
We compute the corresponding component of the spectrum, and
exhibit the eigenvalues explicitly.   The other eigenstates are related
to a generalization of the spin networks, and their eigenvalues can be
computed by diagonalizing finite dimensional matrices.  Furthermore,
we show that the eigenstates of the volume diagonalize also the area
operator.  We argue that the spectra of volume and area determined
here can be considered as predictions of the loop-representation
formulation of quantum gravity on the outcomes of (hypothetical)
Planck-scale sensitive measurements of
the geometry of space.

\vskip .3cm
\noindent
 \vfill
${}^* $ rovelli@vms.cis.pitt.edu,
$\ \ {}^\dagger$  smolin@phys.psu.edu
\eject

\section{Introduction}

In spite of recent progress, research in quantum gravity\cite{cris}
has produced few precise physical predictions, against which
the theory might be, at least ion principle, experimentally
tested.
In this paper, we show that, under certain assumptions,
predictions for the spectra of certain geometric quantities can be
derived from the quantum theory of gravity
in the formulation based on the loop
representation\cite{lp1,lp4,lp5,lp6}\footnote{See also
\cite{math,math3,gambini}. For detailed, but not up to date, reviews
see \cite{cr-review,aa-review,ls-review}.}.  More precisely,
we show that the operators that correspond to the
measurements of the volumes and areas of appropriately defined
regions can be defined in this formulation of quantum gravity.
We are able to diagonalize these operators, and we find that the
spectra fall into certain discrete series, some of which we are
able to explicitly determine.  Finally, we argue  that these
spectra admit an interpretation as physical predictions of the
loop-representation approach to quantum gravity.

To establish these or any other physical predictions
from non-perturbative formulations of quantum
gravity, there are
two main sources of difficulties that must be overcome. The
first of these is
the quantum-field-theoretical problem of defining operator products
in the absence of the background geometry on which conventional
regularization schemes rely.  The second is the fact that in a
diffeomorphism-invariant theory physical observables express
correlations amongst degrees of freedom and are therefore
non-trivial functions of
the elementary fields.  Each of these difficulties is addressed
in this paper, and the methods by which this is done are the
main technical underpinnings of our results.  We address the first of
these difficulties by utilizing a recently developed technique for
constructing diffeomorphism-invariant
regularizations of certain operator products\cite{ls-review}.
This technique allows us to
construct area\cite{lp4} and volume operators and to compute their
spectra.  We address the second difficulty by considering
physically meaningful observables which are obtained by
using dynamical matter fields to construct
spacetime reference frames, an old idea recently discussed in
\cite{whatiso,mc,mc2,matterrefl,ls-brill,karel}. This strategy,
combined with the non-dynamical character of quantities we are
considering, allows us to endow the spectra we compute with a
physical interpretation\cite{mc,mc2}.

The volume $V(R)$ of a three-dimensional spatial region $R$
is determined by the spacetime metric; in the context of quantum
relativistic gravitation, therefore, the volume expresses an
observable function of the quantum gravitational field.
The main work of this paper is the construction of the operator
that represents the quantum
observable $\hat{V}(R)$ in the loop representation, and the
study its
associated spectral problem.

A key point, which underlies everything that is done here is that
both the volume and area are non-local functions of the local
metric or frame fields.
Local quantities tend to be
divergent or ill-defined in non-trivial generally covariant quantum
field theories \cite{lp4,lp5,ls-review}, a fact likely to be related to
the non-renormalizability of the perturbative series.  The
regularization technique we employ for constructing the volume
operator \cite{ls-review} exploits the non-locality of the volume in
order to circumvent this non-renormalizability of local
quantities.  An unphysical
background metric is introduced for the sake of the
regularization, but the operator turns out to be independent of
this background, in the limit in
which the cut-off is removed.  Furthermore, as a consequence,
for reasons that have been discussed
previously\cite{lp4,lp5,ls-review}
the operator is finite and diffeomorphism invariant.  We may note
that this strategy
differs significantly from that involved in the definition of
operator products in conventional quantum field theories on
fixed background fields.  We expect that this approach to
regularization
might be of general
interest for the construction of non-trivial diffeomorphism
invariant quantum field theories in the several different contexts in
which they arise.\footnote{For various
recent perspectives on diffeomorphism invariant quantum field
theories see \cite{atiyah,baez}, and reference therein.}

Our main result is that we find that the spectrum of the
volume is discrete and that eigenstates and eigenvalues can be
expressed in terms of the spin network calculus, a
technology developed
many years ago by Roger Penrose with a quite different purpose
\cite{penrose}.  We compute one component of the spectrum
explicitly (eq.(1) below) and define a straightforward strategy for
computing the rest of it.

Penrose introduced the spin networks as an attempt to construct a
quantum mechanical description of the geometry of space.  A spin
network is a trivalent graph $\Gamma$ (a graph in which each node
joins three links), in which a positive integer $p_l$  (a ``color"), is
associated to every link $l$ of the graph (see Figure 1), with certain
restrictions on the colors at each node.
\begin{figure}[t]
\epsffile{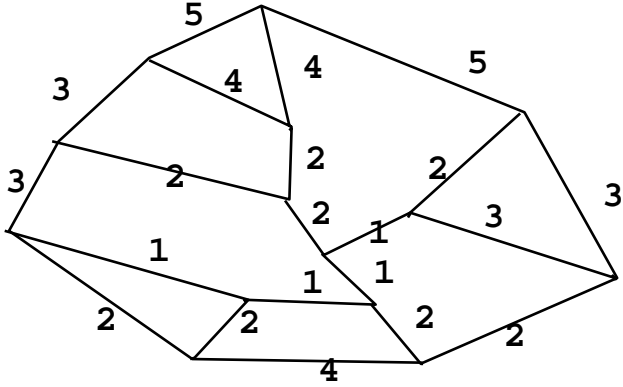}
\caption{A spin network}
\end{figure}
As shown in \cite{spinnet}, one may associate a quantum state of
the gravitational field to each spin network;  these states are
interesting because they can be used to define a basis of linearly
independent loop states.\footnote{Arbitrary loop states
corresponding to different loops are generally not linearly
independent, due to the spinor, or $SU(2)$-Mandelstam,
identities \cite{lp1}.} Here we show that the quantum states of
the gravitational field defined by the trivalent spin networks are
eigenstate of the volume operator.  The eigenvalues are given as
follows.  Let $p_i, q_i, r_i$ be the colors of the links adjacent to the
$i$-th node of the spin network, and  let $a_i, b_i, c_i$ be defined by
$p_i = a_i + b_i,\  q_i=b_i + c_i,\  r_i=c_i + a_i$.
(The $a_i, b_i, c_i$
are always integer due to the constraints on the coloring, and have a
simple geometrical interpretation.)
We then find that the eigenvalue of the operator that
corresponds to the volume of a region $R$
is
\begin{equation}
	V = {1\over 4}\ l_{P}^3 \ \sum_{i}
	\sqrt{a_i b_i c_i + a_i b_i + a_i c_i + b_i c_i},
\end{equation}
where $l_P$ is the Planck length and the sum runs over all the nodes
$i$ contained in $R$.  This formula provides a component of the
physical spectrum of the quantum volume.

Furthermore, we
show here that the spin network states also diagonalize  the
operator $\hat{A}[{\cal S}]$, introduced in \cite{lp4}, which
measures the area of the surface $\cal S$.  (The two
operators obviously commute.)  We complete the
diagonalization of the area, and we compute its full spectrum, which
is
\begin{equation}
	A = {1\over 2}\ l_{P}^2\ \sum_{l} \sqrt{p_l^2 + 2p_l}
\end{equation}
where $l$ labels the links of the spin network that cross the
surface.

Given these results, we may ask whether there
is there a relation between the spectra (1-2), and the volumes and
areas that we physically observe.  Physical information, including
the information about geometry, resides in
the diffeomorphism invariant
observables\footnote{For a discussion on the way in which
purely gravitational diffeomorphism invariant observables (and
loop-kind observablesin particular) code the full physical
information on spacetime geometry, see \cite{carlip}, where the 2+1
case, which is not plagued by the field-theoretical difficulties of
the 3+1 case, is discussed.}.   If we assume that the universe is
spatially compact, then the volume of a 3-dimensional spatial slice
$\Sigma$ of the universe is obviously invariant under 3-dimensional
diffeomorphisms of $\Sigma$.   If $\Sigma$ is determined by
the value of a physical field which is part of the dynamical theory
considered (for instance if it is defined as a constant value surface
of a scalar field), then this volume is also 4-dimensional
diffeomorphism invariant, and hence physically
meaningful.\footnote{Under certain conditions the scalar field may
play the role of a clock, and evolution with respect to such a clock
field can be considered \cite{lp5,ls-brill,karel};
the volume operator can then be seen as a time dependent
Heisenberg
operator that represents the volume of the universe at a certain
clock time.}  In addition, if the theory includes also physical fields
that determine finite spatial regions
\cite{whatiso,mc,mc2,matterrefl,ls-brill,karel},
then the volumes of these regions too are 4-dimensional
diffeomorphism invariant observables.  The same holds for the area
of surfaces  defined by physical fields.  Thus, in the presence of
dynamical matter, area and volume of matter-determined surfaces
are diffeomorphism invariant quantities. A very important
observation is that {\it any\ } actual volume or area measurement
that we may concretely perform defines such a 4-dimensional
diffeomorphism invariant observable, because any region whose
volume or area we measure is determined by physical fields and, in
the context of relativistic gravitation, these fields are necessarily
coupled to the gravitational field \cite{whatiso}.  We shall argue
below that the physical area and volume operators corresponding to
those measurements in the context of a gravity-matter theory are
the same mathematical operators as the pure gravity ones
constructed here. It follows that the spectra are equal; and
therefore the spectra we derive here can be interpreted as physical
predictions from the quantum theory of gravity in the loop
representation:
\begin{itemize}
\item \em If one measured the volume of a physical region or the
area of a physical surface with Planck scale accuracy, one would
find that any measurement's result falls into the discrete spectra
given here [as in eqs.(1-2)].
\end{itemize}

In section 2 we construct the regularization procedure that yields
the volume operator.  This construction corrects certain difficulties
of a previous definition \cite{lp5,ls-review}.  In section 3 we
compute the action of the operator on spin network states, and we
calculate its spectrum.  To this purpose, we make use of the results
and the technology developed in \cite{spinnet}.  This technology
simplifies the geometrical-combinatorial calculus on loop space,
allowing the full computation of the action of the three-hands loop
operator which enters the definition of the volume.  In section 4 we
study the area operator and derive its full spectrum.  In section 5 we
discuss the physical interpretation of the spectra.  In section 6 we
discuss the general result obtained.

\section{Construction of the volume operator}

Let us consider a three dimensional region $R$.  The volume of this
region is given by
\begin{equation}
	V = \int_R \! d^3x \, \sqrt{\det g},
	\label{int}
\end{equation}
where $g$ is the three-dimensional metric of space.  In terms of
Ashtekar variables \cite{abhay}, the volume element can be written
as
\begin{equation}
	\det g =\left |  \det \tilde{E}  \right | ={1 \over 3!}
	\left | \epsilon_{abc} \epsilon_{ijk} \tilde{E}^{ai}
	\tilde{E}^{bj}\tilde{E}^{ck}
	\right | .
\end{equation}
Indices $a, b, ...=1,2,3$ label tangent-space vector components and
indices $i, j, ...=1,2,3$ internal space components; we refer to
\cite{cr-review,aa-review,ls-review} for notation and background.

If we try to construct the local quantum field operator
corresponding to the volume element $\sqrt{\det g }$, we encounter
the usual problem of making sense of a non linear function of
operator-valued distributions.  Here,  $\sqrt{\det g}$ involves the
square root of the cube of the field operator ${\tilde{E}}^{ai}(\vec
x)$.  As in any quantum field theory, one may try to circumvent
(undefined) products of operator-valued distributions by means of a
regularization procedure. In the present context, however, the
necessity of preserving diffeomorphism covariance frustrates the
attempts to renormalize $\sqrt{\det g}$, as we shall discuss in a
moment.   The strategy that works is to sidestep the
definition of the quantum operator $\sqrt{\det g}$ by constructing a
regularization procedure that yields directly the non-local operator
$\hat V$.  This strategy was successfully introduced in
\cite{lp4} for the area operator, and is based on the (non-local)
loop operators of the extended loop algebra \cite{lp1}.

Consider the three-indices (``three-hands") loop observable
\cite{lp1}
\begin{equation}
	T^{abc} [\alpha ](s,t,r) = Tr \left\{
	\tilde{E}^{a}(\alpha (s))\, U_\alpha (s,t)\,
	\tilde{E}^{b}(\alpha (t))\, U_\alpha (t,r)\,
	\tilde{E}^{c}(\alpha (r))\, U_\alpha (r,s)
	\right\} ,
\end{equation}
where $U_\alpha(s,t)$ is the parallel propagator of the
Ashtekar connection along the loop $\alpha$ from the value $s$ to
the value $t$ of the loop parameter, and $\tilde{E}^a = 4 \tilde{E}^{ai}
\tau^i$, where the generators of $SU(2)$ are given by $\tau^i=-
{\imath \over 2} \sigma^i$, and $\sigma^i$ are the Pauli matrices.
Using the identity
\begin{equation}
	Tr(\tau^i\tau^j \tau^k ) = -{1 \over 4} \epsilon^{ijk},
\end{equation}
we notice that in the limit in which the loop $\alpha $ shrinks to a
point $x$ we have
\begin{equation}
	\lim_{\alpha \rightarrow x}\epsilon_{abc}
	T^{abc}[\alpha](s,t,r)
	= - 16\ 3! \ \det \tilde{E} .
	\label{16}
\end{equation}
Trying to implement this limit in the quantum theory yields an
uncontrolable divergence: in fact, if we renormalize
the operator so that the limit remains finite, we find that the free
renormalization ``constant" produced by the renormalization
transforms under diffeomorphisms as a scalar density
\cite{lp4}.  This fact amounts to an infinite-fold
renormalization ambiguity, because there is no preferred scalar
density field on a differentiable manifold without background metric
structure.  The local quantum operator $\sqrt{\det g}$ is thus
ill-defined in the theory.  Notice
that this difficulty is a direct consequence of the fact that the
quantum field theory is to be defined on a differential manifold,
namely on a much weaker structure than the metric manifolds on
which quantum field theories are conventionally defined. We are
indeed confronting here the very core of the problem of quantum
gravity: making sense of a general covariant quantum field theory.

As we mentioned already, we deal with this difficulty by
representing directly the volume of a {\it finite\,} region via a
limiting procedure that does not require the (ill-defined) local
operator $\sqrt{\det g}$ as an intermediate step.  To this aim, let us
fix an arbitrary (unphysical) flat metric $g^0$ in $R$, and let us
partition the region $R$ into cubic boxes (cubic with respect to
$g^0$), whose sides have length $L$ (with respect to $g^0$).  We
shall later verify that our results are independent from $g^0$ and
from the partition chosen.   We label the boxes with an index $I$.
Clearly
\begin{equation}
	V =\sum_I V_I ,
\end{equation}
where $V_I$ is the volume of the $I$-th  box.  If $L$ is small enough,
by the very definition of the Riemann integral in (\ref{int}), $V_I$ is
approximated by $L^3 \sqrt{\det g(x_I)}$, where $x_I$ is any point
in
the $I$-th box, and
\begin{equation}
	V=\lim_{L \rightarrow 0} \sum_I L^3
	\sqrt{ \left | \det \tilde{E}(x_I) \right | } .
\end{equation}

We now replace $\det \tilde{E}(x_I)$ by a non-local
quantity that approximates it for small $L$.  In the quantum theory
this non-local quantity will remain finite for any finite $L$.
Consider one particular box, and let $\partial I$ indicate its
boundary, namely the union of the six faces of the cube, oriented
outwards.   Consider three arbitrary points $\sigma,\tau$ and $\rho$
on this boundary.   Given these three points, let
$\alpha_{\sigma\tau\rho}$ be the triangular loop formed by the
three straight (in
the
metric $g^0$) segments that connect the three points $\sigma, \tau$
and $\rho$.    We define
\begin{equation}
	{\cal W}_I \equiv
	\int_{\partial I}\! d^2  \sigma
	\int_{\partial I}\! d^2  \tau
	\int_{\partial I}\! d^2  \rho \
	\left |
	     n_a ( \sigma) n_b ( \tau) n_c ( \rho)\,
	     T^{abc}[\alpha_{ \sigma\tau\rho}](s, t, r)
	\right | ,
	\label{w}
\end{equation}
where $n_a$ is the cube's boundary outward-pointing normal vector
(more precisely, it is the one-form tangent to the boundary
of the box) and the integration measure is the area element induced
by  $g^0$. Furthermore, $s, t$ and $r$ are the values of the
$\alpha_{\sigma\tau\rho}$ loop-parameter at the three vertices of
the triangle, namely $\alpha_{\sigma\tau\rho}(s)=\sigma, \
\alpha_{\sigma\tau\rho}(t)=\tau,\
\alpha_{\sigma\tau\rho}(r)=\rho$: in other words,
the hands of the loop observable sit on the box's boundary at the
corners of the triangular loop.  See Figure 2.

\begin{figure}[t]
\epsffile{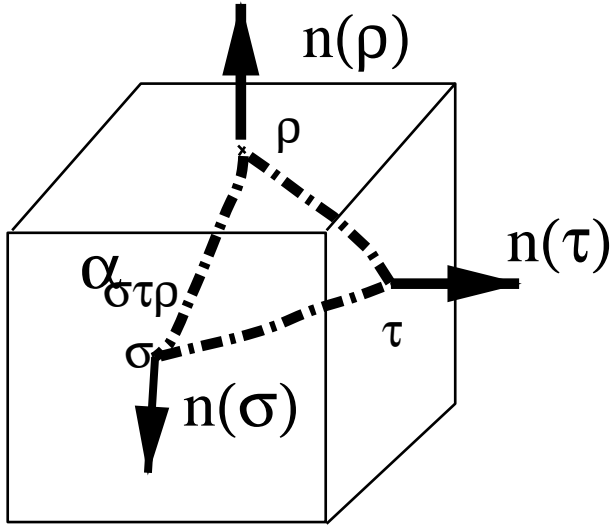}
\caption{The $\hat T^{abc}$ operator}
\end{figure}

Then it is easy to verify that
\begin{equation}
	{\cal W}_I = 2^{7} 3!\, L^6\ \left |
	\det \tilde{E} \right |
	+ O(L^7).
\end{equation}
The factor $2^7$ comes from the $16$ in equation (\ref{16}) and
from the fact that the triple integral on orthogonal faces of the cube
yields a $(2L^2)^3$ term, since there are 2 faces of area $L^2$ for
every orientation.  The quantity $(2^7 3! L^6)^{-1}{\cal W}_I$ is thus
a non-local quantity that approximates $\det \tilde{E}(x_I)$ for
small $L$.  Bringing everything together, we can write the total
volume as the limit
\begin{equation}
	V = \lim_{L \rightarrow 0}  \sum_I
	\sqrt{
		{\scriptscriptstyle{1\over 2^7 3!} } \
		{\cal W}_I
		} .
	\label{pippo}
\end{equation}
Notice that this expression is a
$SU(2)$-gauge-invariant point splitting of the cubic product in
the definition of the volume.  Equations (\ref{w}) and (\ref{pippo})
may not provide the
easiest way of calculating a volume classically,
but, as we shall see, the sum
in (\ref{pippo}) can be promoted to an operator which is
well-defined in the limit, in spite of the fact that the local volume
element
is ill-defined.  Thus, we define the quantum operator corresponding
to the physical volume of the region $R$ as
\begin{equation}
	\hat{V} = \lim_{L \rightarrow 0} \sum_I
	\sqrt{{\scriptscriptstyle{1 \over 2^7 3!}}\ \hat{\cal W}_I} ,
	\label{hatv}
	\label{qpippo}
\end{equation}
where $\hat{\cal W}_I$ is given by (\ref{w}) with $T^{abc}$ replaced
by the corresponding operator $\hat T^{abc}$.  What is going to
happen is that the quantum $(\det\tilde{E}(x_I))$ diverges as $L^{-
6}$ (and thus in a badly background dependent way, because there
is
no covariant meaning in giving the same size to all boxes), therefore
$\hat{\cal W}_I$ is finite in the limit; in addition, however, all but
a finite number of terms in the sum (\ref{qpippo}) will vanish in the
limit, so that (\ref{qpippo}) turns out to be finite.

The definition (\ref{qpippo}) of the quantum operator $\hat V$
requires us to specify the topology in which the limit is taken
\cite{ls-review}.  We take here the weak operator topology
(convergence in the "matrix elements" $\langle\alpha|\hat
V|\psi\rangle$), and we define the operator on the domain $\cal D$
formed by the loop functionals $\psi(\alpha) = \langle\alpha |
\psi\rangle $ continuous in the loop space topology naturally induced
by the manifold topology. (For an arbitrary smooth metric $d$ on the
manifold, the $\epsilon$-neigborhood of $\alpha$ is formed by the
loops $\beta$ that admit a parametrization in which $d(\beta(s)-
\alpha(s)) < \epsilon$ for all $s$.)  Once defined on this domain, the
operator can then be extended to its maximal domain ${\cal D}_{{\rm
max}}$.   This is a natural choice\footnote{As is well known, in
promoting composite observables to quantum operators there might
remain a degree of arbitrariness not entirely constrained by
symmetry or consistency: quantization is an inverse problem, which
may have more than one solution. The task here is to understand if
there is {\it a\,} consistent quantum theory of gravity and to unravel
its physical consequences, not to find {\it the\,} incontrovertibly
correct one out of pure thought.}, but we should point out that the
issue is delicate, because the diffeomorphism invariant states are
in ${\cal D}_{{\rm max}}$, but not in $\cal D$.
The mathematical-physics technology which is being developed by
Ashtekar, Isham and collaborators \cite{math,math3} should allow
to clarify these subtelties; note for instance that there are other
relevant topologies in state space, as the norm topologies defined by
the kinematical or diffeomorphism-invariant Hilbert structures
\cite{math,math3}, which are inequivalent to the one in which the
limit (\ref{qpippo}) is taken. See section 4.2 of \cite{ls-review} for
a discussion of this point.

For completeness, let us recall from \cite{lp1} the definition of the
three-hand operator $\hat{T}^{abc}$:
\begin{eqnarray}
	\langle \beta| \hat{T}^{abc}[\alpha ](s,t,r)
	&=& l_{P}^6\
	\Delta^a [\beta, \alpha (s)]
	\Delta^b [\beta, \alpha (t)]
	\Delta^c [\beta, \alpha (r)]
	\nonumber  \\
	&&
	\times \sum_{j=1,8} (-1)^{r(j)} \langle (\alpha \#_{str}
	\beta )^j |.
	\label{t3}
\end{eqnarray}
The distributional factor $\Delta^a [\beta , x ]$ is
\begin{equation}
	\Delta^a[\beta, x ]
	=\int\! du\ \dot{\beta}^a (u)\ \delta^3 (\beta (u), x)
\end{equation}
and $j$ labels the eight different loops $(\alpha \#_{str} \beta
)^j$ obtained combining $\alpha $ and $\beta $ by breaking and
rejoining them in all possible ways at the three points of
intersection $s, t, r$, where the operator acts (which we call
``grasps").  The sign factor $r(j)$ is defined in \cite{lp1} as the
number of times that the orientation of loops' segments (coming
from an arbitrary orientation of $\alpha$ and $\beta$) must be
switched to yield a consistent routing; this cumbersome way of
keeping track of the signs is superseded by the much
simpler Penrose notation introduced in \cite{spinnet}, which we
shall recall and use below.

In the equation (\ref{t3}) the conventional bra notation
of the loop representation is used (the loop functionals notation is
obtained by contracting with a ket on the right: $\langle\alpha|\hat
O|\psi\rangle = \hat O\psi(\alpha)$.)   In this paper we shall find the
left eigenvectors of $\hat V$.  To the extent the theory is consistent
(which is not proven), $\hat V$ must be hermitian and left and right
spectra should coincide.  Also, for notational simplicity, we shall
from now on reverse the convention and write all the loop states as
kets (so equation (\ref{t3}) becomes $\hat{T}^{abc}[\alpha ](s,t,r)
|\beta\rangle = ... $).

\section{Action of the volume operator}

Our task is to compute the action of the operator $\hat V$, given in
eq.(\ref{hatv}), on arbitrary quantum states.  To this aim, we
chose to work in a particular basis: the spin network basis
\cite{spinnet}.   As we will see, this basis essentially diagonalises
the  operators that we are considering; this will allows us to deal
with the absolute value and the square route, which otherwise could
have been sources of substantial difficulties.

We recall here the main elements of the definition of the spin
network states \cite{spinnet}.  A set of $n$  fully overlapping loop
segments is denoted as an $n$-rope, or a rope of degree $n$.  An
intersection
between loops is called $k$-valent if $k$ ropes (of any degree)
emerge from it, and a loop state $|\gamma\rangle$ is called
$k$-valent if $\gamma$ contains intersections of valence at most
$k$.  We will begin
by studying the action of $\hat V$ on trivalent states, and consider
higher order intersections at the end of this section.  A basis for the
trivalent loop states is given by the spin network states.  An
(imbedded) spin network $(\Gamma, p_l)$, is a trivalent graph
$\Gamma$, imbedded in the spatial manifold $\Sigma$ (which we
take here compact and with fixed topology, say $S^3$), in which each
link $l$ is colored by a positive integer $p_l$, namely a positive
integer is assigned to every link.  At each node $i$ of the graph, the
colors of the three links adjacent to the node satisfy two relations:
their total sum is even, and none is larger than the sum of the other
two.   A spin network quantum state
$|\Gamma, p_l\rangle$ is obtained by a given spin network
$(\Gamma,
p_l)$ as follows:  First, replace every link $l$ with a rope of
degree $p_l$.  Then, at each node join all the segments of the three
adjacent ropes pairwise, in such a way that each segment is joined
with a segment of one of the other two ropes.  The conditions on the
coloring make the matching always possible. Since all segments' end
are joined, we obtain in this way a (multiple) loop
$\gamma^{(\Gamma, p_l)}$.  Consider the corresponding loop state
$|\gamma^{(\Gamma, p_l)}\rangle$.  Then, consider all
possible permutations of the joining of all segments along every
rope, so that the original choice of pairing becomes
irrelevant.  In this way, one obtains a family of $M = \prod_l p_l !$
(multiple) loops, which we denote as $\gamma^{(\Gamma, p_l)}_m,
m=1, ..., M$, or, if the context is clear, simply as
$\gamma_m$.  Each one of
these loops have support $\Gamma$.  The quantum state  $|\Gamma,
p_l\rangle$ associated to the spin network $(\Gamma, p_l)$ is
defined as
\begin{eqnarray}
	 |\Gamma, p_l\rangle & = & \sum_m \
	\epsilon_m \
	|\gamma^{(\Gamma, p_l)}_m\rangle.
	\label{ds}
\\
	\epsilon_m & = & (-1)^{(p_m+n_m)}
\end{eqnarray}
where $n_m$ is the number of single loops in the multiple loop
$\gamma^{(\Gamma, p_l)}_m$ and $p_m$ is the parity of the
segments' permutation defining $\gamma^{(\Gamma, p_l)}_m$ (the
overall sign, irrelevant for what follows, is not determined by this
definition; see \cite{spinnet}).

One may think of a spin network state as the loop transform of the
connection-representation state obtained by replacing every link
colored $p$ by parallel propagators of the Ashtekar connection,
taken in the spin $p/2$ representation, and then contracting the
three parallel propagator matrices at each node in
the unique $SU(2)$-invariant way, namely using the $6j$ symbols.
The conditions on the colorings at each node then reflect the algebra
of the $SU(2)$ representations, namely the angular momentum
addition rules.

Consider the $i$-th node of a given spin network, and let $p$, $q$
and $r$ be the colors of the three adjacent links.  We define $a_i,
b_i, c_i$ by
\begin{equation}
	p = a_i + b_i, \ \ \
	q = b_i + c_i, \ \ \
	r = c_i + a_i.
	\label{abc}
\end{equation}
When clear from the context, we will drop the suffix $i$ and write
just $a, b, c$.  In
constructing the loop state from the spin network, $a_i$ is the
number of individual loops that are routed through the $i$-th node
from the link colored $p$ to the link colored $r$, and so on, as shown
by Figure 3.
\begin{figure}[t]
\epsffile{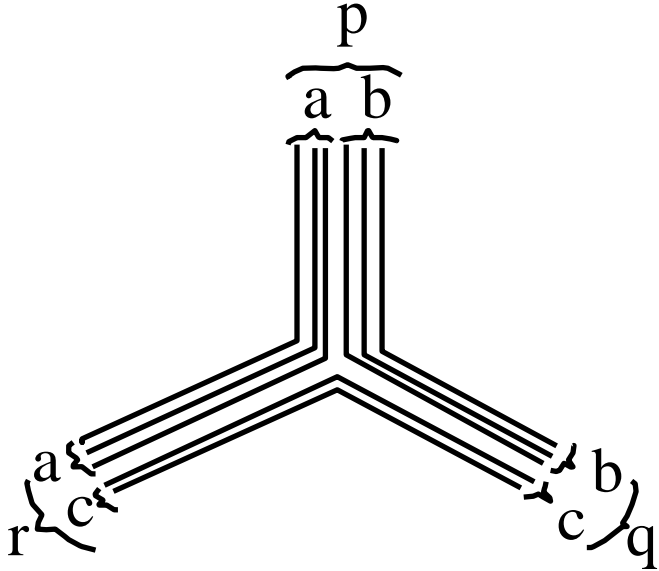}
\caption{The integers $a, b$ and $c$}
\end{figure}
The numbers $a_i, b_i, c_i$ provide a way of
coding the coloring, but they are not independent:  in a spin network
with $N$ nodes the $3 N$ numbers $a_i, b_i, c_i$ are related to each
other because if a link (of color $p$) connects the $i$-th and $j$-th
node  then $p = a_i+b_i = a_j+b_j$, and so on for each link of the
spin network; there is one linear relation of this kind for every link.
We shall use also the notation
$(\Gamma, a_i b_i c_i)$ for a spin network, and the notation
$|\Gamma, a_i b_i c_i\rangle$ for the corresponding quantum state.
For later convenience, we introduce also the following notation for
the points along the ropes and their tangent:  we parametrize each
link $l$ of the graph with a parameter $s$, and we denote the points
of the links as $l(s)$ and their tangents as  $\dot
l^a(s)$.  We will use $l(s)$ and $\dot l^a(s)$ in
place of more cumbersome expressions such as $\gamma^{(\Gamma,
p_l)}_m(s)$ and $(\dot\gamma^{(\Gamma, p_l)}_m)^a(s)$ when the
context is clear.

We want to compute the action of the volume operator on a
(trivalent) spin network state.  Using
the definitions (\ref{hatv}), and (\ref{w}), we have
\begin{eqnarray}
	\hat{V}\ |\Gamma, a_i, b_i, c_i\rangle &=&
	\lim_{L \rightarrow 0} \sum_I
	\sqrt{{\scriptscriptstyle{1 \over 2^7 3!}}\ \hat{\cal W}_I}\  \
	|\Gamma, a_i b_i c_i\rangle
\nonumber \\
	 &=& \lim_{L \rightarrow 0} \sum_I
	\left({\scriptscriptstyle{1 \over 2^7 3!}}
	\int_{\partial I\times\partial I\times\partial I} d^2  \sigma
	\,  d^2  \tau \,  d^2  \rho \right.
\nonumber \\
	&&   \times \left.
	\left |
	   n_a ( \sigma) n_b ( \tau) n_c ( \rho)\
       \hat T^{abc}[\alpha_{ \sigma\tau\rho}](s, t, r)
	\right | \  \right)^{1/2}
\nonumber \\
	&&  \times  |\Gamma, a_i b_i c_i\rangle .
	\label{lungo}
\end{eqnarray}

Let us fix a box $I$ and three points $ \sigma,\tau,\rho$
on its boundary, and let us consider the action of the operator valued
distribution $\hat T^{abc}[\alpha_{ \sigma \tau
 \rho}](s,t,r)$ on $|\Gamma, a_i b_i c_i\rangle$.  Using the
definition (\ref{ds}) of a spin network state and the
definition (\ref{t3}) of the three-hands loop operator, we have
\begin{eqnarray}
	&& \hat T^{abc}[\alpha_{ \sigma\tau\rho}](s,t,r)
	\, |\Gamma, a_ib_ic_i\rangle  =
	\hat T^{abc}[\alpha_{ \sigma\tau\rho}](s,t,r) \
	\sum_m  \epsilon_m |\gamma_m\rangle
\nonumber \\
	&&\ = \sum_m  \epsilon_m
	\int ds'\, \dot\gamma_m^a(s')\,
	 \delta^3(\gamma_m(s'), \sigma)\,
	 \int dt'\, \dot\gamma_m^b(t')\, \delta^3
	(\gamma_m(t'), \tau)
\nonumber \\
	&& \ \ \ \times \
	 \int dr'\, \dot\gamma_m^c(r')\,
	\delta^3 (\gamma_m(r'), \rho)\
	\sum_{j=1,8} |(\alpha_{ \sigma\tau\rho}
	\#_{(str)} \gamma_m)^j\rangle
\end{eqnarray}
(here, clearly, $\gamma_m$ means $\gamma^{(\Gamma,  a_i b_i
c_i)}_m$). The right hand side of this equation vanishes
except in the case in
which the three hands of the operator intersect $\Gamma$.
Therefore, we
have a non vanishing result only if the three points $ \sigma,
 \tau,  \rho $ are intersections points between the
boundary of the box and $\Gamma$.   Let us assume this is the case.
 We denote
the three ropes that intersect $ \sigma,  \tau,  \rho $
by $l_\sigma, l_\tau, l_\rho$ respectively (they may coincide), and
their degree  by $p_\sigma, p_\tau$ and $p_\rho$.  The three ropes
are thus formed by $p_\sigma, p_\tau$ and $p_\rho$ segments
respectively, each one of which is seen by one of the hands
 of the operator. Let us
 introduce an index $S$, running from 1 to $p_\sigma$ that labels
the $p_\sigma$
segments in the rope $l_\sigma$, and, similarly, indices
$T=1...p_\tau
$ and
$R=1...p_\rho$ labeling the segments of $l_\tau$ and $l_\rho$
respectively;
and let us denote by $s_S, t_T, r_R$ the values of the loop
parameter corresponding to the intersections
between the box boundary and the $S$-th, $T$-th and $R$-th
segments of the rope $l_\sigma, l_\tau$ and $l_\rho$ respectively.
We have then
\begin{eqnarray}
	&&\hat T^{abc}[\alpha_{ \sigma\tau\rho}](s,t,r)
	\ |\Gamma, a_ib_ic_i\rangle  =
\nonumber \\
	&&  = \! \int ds\, \dot l_\sigma^a(s)\,
	 \delta^3(l_\sigma(s), \sigma)
	\!\int dt\, \dot l_\tau^a(t)\, \delta^3
	(l_\tau(t), \tau)
	\! \int dr\, \dot l_\rho^a(u)\,
	\delta^3 (l_\rho(r), \rho)
\nonumber \\
	&&  \times  |\Psi\rangle,
	\label{due}
\end{eqnarray}
where
\begin{equation}
	|\Psi\rangle = \sum_m \ \epsilon_m\ \sum_S \sum_T \sum_R
 	\sum_{j=1,8} |(\alpha_{ \sigma\tau\rho}
	\#_{(s_S t_T r_R)} \gamma_m)^j\rangle.
\end{equation}
We repeat for clarity: $m$ labels the multiple loop states that form
the spin network state, or, equivalently, labels all the possible
symmetrizations of all the segments along each rope of the net;
$S, T$ and $R$ label the segments in the three ropes that
intersect the box; once fixed one of these
segments in each of the three ropes, $j$ labels the eight outcomes
of the
three graspings of the operator on these three segments.

Now, the key observation is the following.  First, let us exchange
the order of the summations in the last equation by summing over
$m$ first.  Then, consider what happens to $\Psi$ in the limit in
which the box size $L$ goes to zero.    The loop
$(\alpha_{ \sigma \tau  \rho}\#_{(s_St_Tr_R)}
\gamma_m)^j$ is obtained by joining the spin network loop
$\gamma^{(\Gamma, a_ib_ic_i)}_m$ with the triangular loop
$\alpha_{ \sigma  \tau  \rho}$ which has size $L$.  In the
limit in which $L$ is zero the three grasping points $ \sigma,
 \tau,  \rho$ overlap and
$\alpha_{ \sigma\tau\rho}$ is a single-point loop,
located, say, in the point $P_I$.  In this limit, the loop
$(\alpha_{ \sigma\tau\rho}\#_{(s_St_Tr_R)}
\gamma^{(\Gamma, a_ib_ic_i)}_m)^j$ has the same support as the
grasped
loop $\gamma^{(\Gamma,a_ib_ic_i)}_m$, namely $\Gamma$, and
can differ from
$\gamma^{(\Gamma, a_ib_ic_i)}_n$ only by a rearrangement of the
joining of
the segments in the point $P_I$.  But by rearranging the way the
segments are joined, we can only obtain one
of the other loop states of the spin network, say $|\gamma^{\Gamma,
a_ib_ic_i}_{m'}\rangle$ for a certain $m'$, because the spin network
state is the linear combination of {\it all\,} loop states obtained
symmetrizing the ropes in all possible ways.  Using the continuity of
the loops functionals in the domain of definition of $\hat V$ (see the
end of section 3), we have thus necessarilly
\begin{equation}
	\sum_m \ \epsilon_m\
	|(\alpha_{ \sigma\tau\rho}
	\#_{(s_Pt_Qu_R)}
	\gamma^{(\Gamma, a_ib_ic_i)}_m)^j \rangle =
	c(S,T,R,j) |\Gamma, a_ib_ic_i\rangle + O(L).
	\label{tremeno}
\end{equation}
In the next subsection we will verify this result explicitly and
compute the proportionality coefficient $c(S,T,R,m)$.
This reconstruction of the full spin network state in the limit is
the technical reason for which the spin network states diagonalise
the volume operator.  Using this result, we have
\begin{eqnarray}
	|\Psi\rangle &=& \sum_{STR} \sum_{j=1,8}
	c(S,T,R,j) |\Gamma, a_ib_ic_i\rangle + O(L)
\nonumber \\
	& \equiv & C(\Gamma, a_i b_i c_i; l_\sigma
	l_\tau l_\rho ) \  |\Gamma, a_ib_ic_i \rangle + O(L),
	\label{tre}
\end{eqnarray}
where we have explicitly indicated that the proportionality
coefficient depends on the spin network and on the ropes grasped.
Obtaining in the limit a state proportional to the one on which the
operator acts is the result that allows us to navigate through the
computation of the volume operator.

To be more precise, the point $P_I$ can be either on a link or on a
node. (If it is on neither, $C$ is zero.)  If it is on a link, it is
easy to see that the rearrangement of the joining of the segments
has no effect on a spin network, because the segments are
symmetrized along the link anyway.  But the same is true also if
$P_I$ is a node, because there is only one existing combination of
fully symmetrized ropes in a trivalent nodes.  This is a central
property of the spin networks, which corresponds to the fact that
there is a unique invariant way of combining three irreducible
representations of $SU(2)$.

We will compute the coefficient of proportionality $C(\Gamma, a_i,
b_i, c_i;  l_\sigma,l_\tau,l_\rho)$ in the next subsection.  For the
moment, we use (\ref{due}) and (\ref{tre}) to write (\ref{lungo}) as
\begin{eqnarray}
	&& \hat{V}\ |\Gamma, a_i b_i c_i\rangle =
	 \lim_{L \rightarrow 0} \sum_I
	\left( {\scriptscriptstyle{1 \over 2^7 3!} }
	\int_{\partial I\times\partial I\times\partial I}
	d^2  \sigma\,  d^2  \tau\, d^2  \rho \
	\right.
\nonumber \\
	 &&\times
	| n_a ( \sigma) n_b ( \tau) n_c ( \rho)\ \
	\sum_{ l_\sigma l_\tau l_\rho}
\nonumber \\
	 &&\times
	\int ds\, l_\sigma^a(s)\, \delta^3 (l_\sigma(s), \sigma)
	\int dt\, l_\tau^a(t)\, \delta^3 (l_\tau(t), \tau)
	\int dr\, l_\rho^a(r)\, \delta^3 (l_\rho(r), \rho)
\nonumber \\
	 && \times
	 \left. C(\Gamma, a_i, b_i, c_i;  l_\sigma l_\tau l_\rho)
	| \right)^{-1/2}\ \   |\Gamma, a_i b_i c_i\rangle.
	\label{quattro}
\end{eqnarray}
We have taken a delicate step here, by taking the limit of a term
inside the integration. \footnote{In a rigorous treatement, one
should check the validity of this step by keeping track of subleading
terms in $L$ and verifying that they remain of lower order also after
the integration.  This is not easy, since the operator is not diagonal
beyond $L=0$, which makes the computation of its square root more
difficult.}

Due to the delta functions, we can take the one-dimensional
integrals outside the absolute value.  Then, we use the well known
fact that for every two dimensional surface $\Sigma $ and every
loop $\beta $
\begin{equation}
	\int_\Sigma d^2 \sigma \int ds\ \  n_a (\sigma ) \dot{\beta}^a
	(s) \ \delta^3 (\sigma, \beta (s) ) = I[\Sigma, \beta ]
\end{equation}
is just the (oriented) intersection number between the surface and
the loop.  This simple relation plays a key role in
loop-representation regularization techniques, because it allows the
action of the operators to be expressed in topological terms
(intersection numbers), which then permits diffeomorphism
invariance to be restored.\cite{ls-review}   Since the absolute value
of a product of three numbers each of which is either 1 or -1 can
only be the unit, we have immediately that
\begin{equation}
	\hat{V}|\Gamma, a_i, b_i, c_i\rangle = \lim_{L \rightarrow 0}
\sum_I
	\sum_{l_\sigma l_\tau l_\rho}
	\sqrt{ {2^{\scriptscriptstyle -7}}
	\left | C(\Gamma, a_i, b_i, c_i; l_\sigma l_\tau l_\rho )
	\right | }
	\  |\Gamma, a_i b_i c_i\rangle
	\label{cinque}
\end{equation}
where the sum $\sum_{l_\sigma l_\tau l_\rho}$
runs over all triples of ropes
intersecting the boundary of the box $I$.   Note that the factor
$1/3!$ has
been absorbed by the fact that there are $3!$ terms coming from the
integrals for each given triple of intersections.

There is an important qualification to be added to the meaning of the
sum $\sum_{l_1l_2l_3}$:  the sum runs only over triples of
{\it distinct\ } intersections; namely the three intersections
between ropes and boundary must lay in three di\-stinct points of
the boundary of the box in order to give a non vanishing
contribution.
The reason is the following. In the derivation of (\ref{cinque}) from
(\ref{quattro}), every triple of intersections gives rise to 3! terms,
obtained by switching the locations of $ \sigma,  \tau $ and $  \rho$.
These terms are equal in magnitude, but have alternating signs,
because $\hat T^{abc}$ changes sign by reversing the order of the
hands. If the three
intersection points are disjoint, then each one of the 3! terms comes
from a different integration point of the triple (six-dimensional)
integration space.  Since we are taking the absolute value first and
then
integrating, the different signs of the terms are irrelevant: we have
to
sum their absolute values.    However, if (at least) two of the
intersection points overlap, then for each of the 3! terms there is a
brother term (obtained by switching the two overlapping
intersections),
which is equal in magnitude, but opposite in sign. The two terms
with
opposite sign arise {\it at the same integration point.}    Thus, we
must
take the absolute value {\it after\ } summing the two terms.
Therefore
they cancel each other.

The important consequence of the fact that only distinct
intersections
enter the sum (\ref{cinque}) is that the action of the operator is
 zero
unless the loop intersects the boundary of the box in at least three
distinct points; a moment of reflection shows that for small enough
$L$,
this can happens only if there is a node
inside the box.  As $L \rightarrow 0$, we reach a point at which each
box
contains at most one node.   Therefore we have the result
that
the sum over the boxes $I$ (which are infinite in the limit) reduces
to a
sum over the nodes $i$ contained in the region $R$ (which
are
finite in the limit).  So we can finally take the limit, and get
\begin{equation}
	\hat{V}\  |\Gamma, a_ib_ic_i\rangle  = \sum_i
	\sum_{l_\sigma l_\tau l_\rho}
	\sqrt{ 2^{\scriptscriptstyle -7}
	\left | C(\Gamma, a_i b_i c_i;l_\sigma l_\tau l_\rho) \right | }
	\ \ |\Gamma, a_i b_i c_i\rangle,
	\label{quasi}
\end{equation}
where now $l_\sigma l_\tau l_\rho$ are the triples of ropes
adjacent to the $i$-th node.
For a trivalent state there is only one
triple of
ropes emerging from every node, and therefore we can write
\begin{equation}
	\hat{V}\ \ |\Gamma, a_ib_ic_i\rangle\ \   = \sum_i
	\sqrt{2^{-7} \left | C(\Gamma, a_i, b_i, c_i; i) \right | }
	\ \ |\Gamma, a_i b_i c_i\rangle.
	\label{trivalent};
\end{equation}
while we will use the expression (\ref{quasi}) for non-trivalent
states.

\subsection{The combinatorics of the routings}

In order to compute $C(\Gamma, a_i b_i c_i; l_\sigma l_\tau
l_\rho)$,
defined in (\ref{tremeno}) and (\ref{tre}), we use the
Penrose notation.  The present exercise
is a demonstration of the power of this formalism.    We must
compute $C$ from
\begin{eqnarray}
	&& \sum_{S=1,p_\sigma} \sum_{T=1,p_\tau}
	\sum_{R=1,p_\tau}
 	\sum_{j=1,8} \sum_m \ \epsilon_m\
	|(\alpha_{ \sigma\tau\rho} \#_{(s_St_Tr_R)}
	\gamma^{(\Gamma, a_ib_ic_i)}_m)^j\rangle =
\nonumber \\
	&& \ \ \ \
	=  C(\Gamma, a_i b_i c_i;l_\sigma l_\tau l_\rho) \ |\Gamma,
	a_ib_ic_i\rangle.
	\label{somma}
\end{eqnarray}
Let the node we are dealing with be the $i$-th one. The sum
includes $8 \times p_\sigma \times p_\tau \times p_\rho $ terms:
we have one of the
three hands of the operator per rope; each hand may grasp any of
the
segments in the rope, giving $p_\sigma \times p_\tau \times
p_\rho$ triple graspings;
and each triple grasping produces $2^3=8$ terms from the two
possible outcomes of each grasp.   The triple graspings produce
different outcomes, according to how the grasped segments are
rooted among themselves through the node.   A short reflection
shows that there are three basic
possibilities that we list in Figure 4 (up to the obvious $2\pi/3$
rotational symmetry).

\begin{figure}[t]
\epsffile{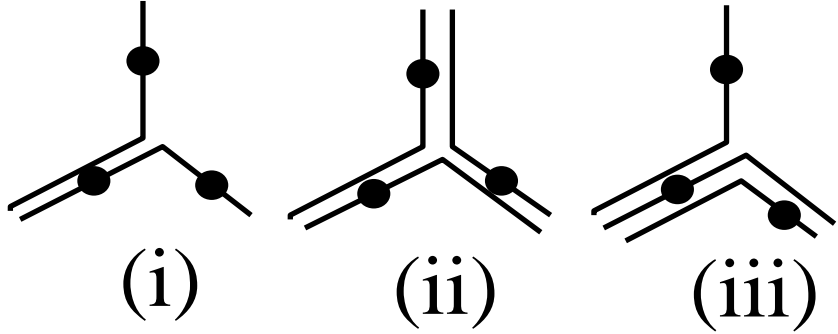}
\caption{The three possible grasps}
\end{figure}
We denote these three possibilities as triple
grasps of the first, second, and third kind.  Counting all possible
instances, the number of (triple) grasps of the first kind (i. $\!\!$ in
the Figure)
that contribute
to the sum (\ref{somma}) is $2abc$; the number of
grasps
of the second kind (ii) is $2(ab+bc+ca)$; and the number of grasps of
the
third kind (iii) is $[a(a-1)(b+c)+b(b-1)(a+c)+c(c-1)(b+a)]$.  Here $a,
b$ and $c$ are defined in terms of $p_\sigma, p_\tau$ and $p_\rho$
as in (\ref{abc}), and we drop
the suffix $i$ for simplicity; we will reintroduce it when needed.
One may verify that the total number of grasps of any kind is
indeed $ p_\sigma \times p_\tau \times p_\rho$.

Let us recall the basic rules of the diagrammatic Penrose calculus
\cite{spinnet}.  The $\hat T^{abc}$ operator is represented by the
figure with three boxes shown in Figure 5; each box standing for a
hand of the operator.
\begin{figure}[t]
\epsffile{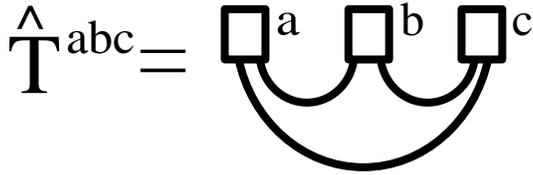}
\caption{The diagrammatic representation of $T^{abc}$.}
\end{figure}

The elementary action of a box is given in Figure 6.

\begin{figure}[t]
\epsffile{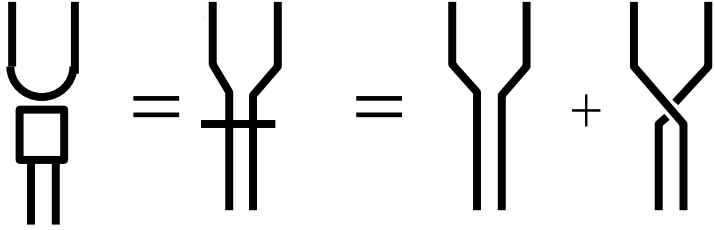}
\caption{The grasp}
\end{figure}

Let us begin by considering the grasp of the first kind
(i).  In diagrammatic notation, this is given in Figure 7, where the
circles labeled $1,2$ and $3$ represent the three ropes emerging
from the node.

\begin{figure}[t]
\epsffile{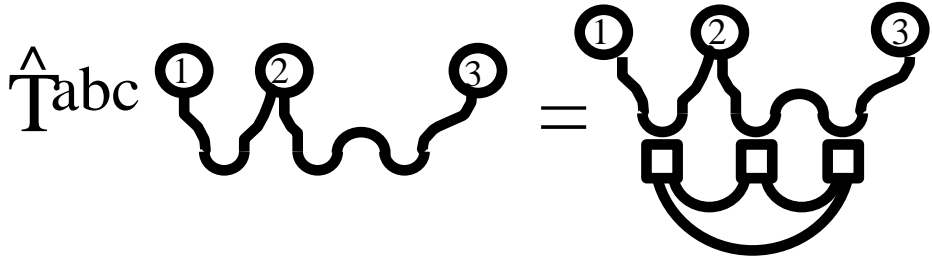}
\caption{The grasp of the $T^{abc}$ on the node of type (i)}
\end{figure}

The result of the grasp is computed using the basic
rule Figure 6, and is given in in Figure 8.

\begin{figure}[t]
\epsffile{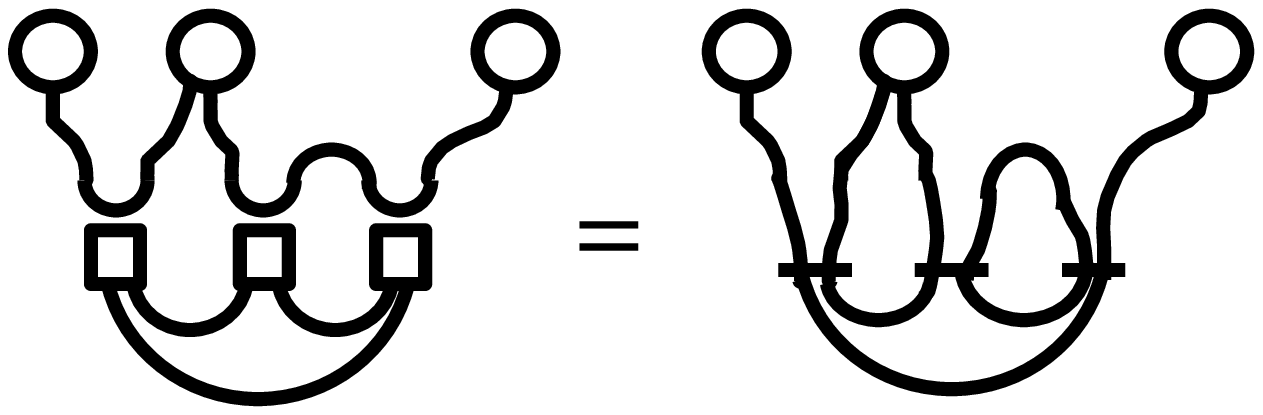}
\caption{Action of the grasp of the $T^{abc}$ on the node of type (ii)}
\end{figure}

Expanding the symmetrizations, we get the $8$ terms of
Figure 9.
\begin{figure}[t]
\epsffile{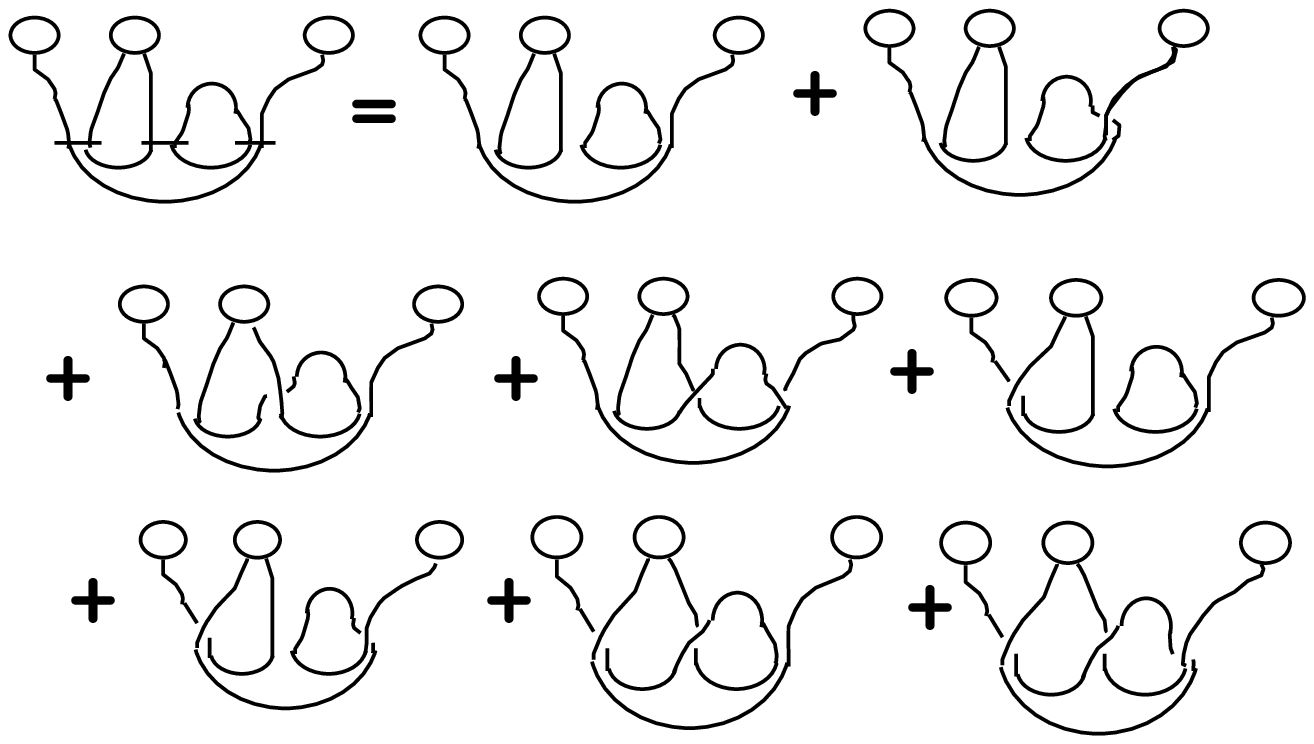}
\caption{Expansion of the symmetrization}
\end{figure}
We see immediately that the first, second and third
of these terms vanish, because all the loops in the rope $2$ are
symmetrized and an identity of the spin calculus is Figure 10.

\begin{figure}[t]
\epsffile{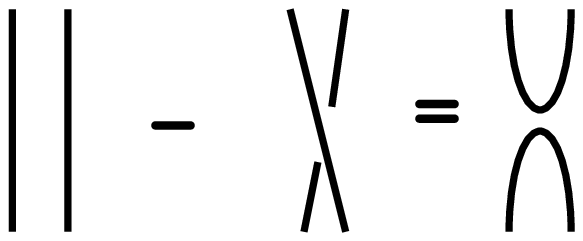}
\caption{The basic spinor identity}
\end{figure}
The remaining terms become, using the sign rules of \cite{spinnet}
those shown in Figure 11, all proportional to the original state.
\begin{figure}[t]
\epsffile{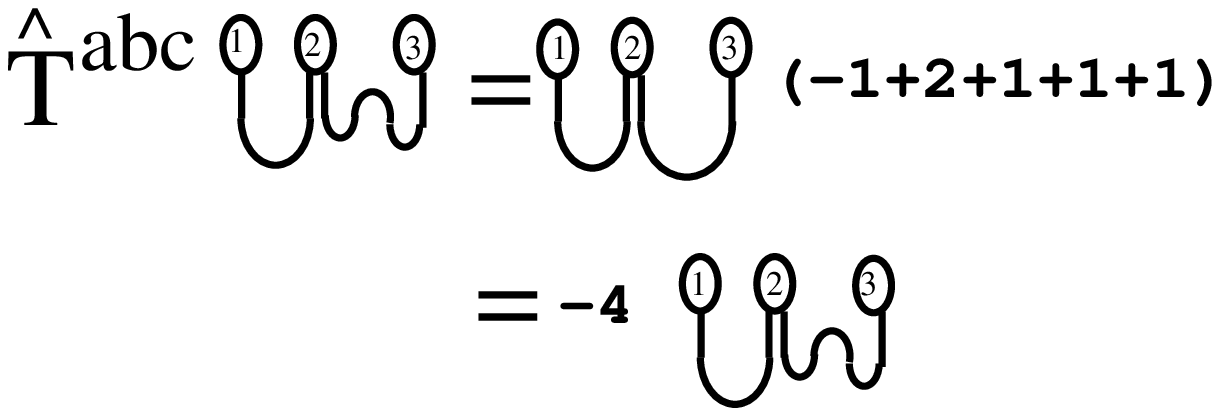}
\caption{The result of the grasp on the node}
\end{figure}

This confirms the expected result that the action of the volume is
diagonal and provide the correct term in the eigenvalue. Recalling
that there are $2(ab+bc+ca)$ terms of this kind in the grasping, we
thus have a contribution $8(ab+bc+ca)$ to $C$ from the graspings of
the second kind.  A completely analogous calculation, which we leave
to the reader as a simple exercise of application of the rules of
\cite{spinnet}, shows that the contribution of the grasp of the first
kind is $8abc$, while the contribution of the grasps of the third kind
vanishes.   Putting all these results together, we have
\begin{eqnarray}
	\lefteqn{\sum_{STR}
 	\sum_{j} \sum_m \epsilon_m
	|(\alpha_{ \sigma\tau\rho} \#_{(s_St_Tr_R)}
	\gamma^{(\Gamma, a_ib_ic_i)}_m)^j\rangle}\nonumber \\
	& & =  2^3 l_{P}^6(abc+ab+bc+ca)\  |\Gamma, a_ib_ic_i\rangle.
\end{eqnarray}
Inserting in (\ref{trivalent}), we have finally
\begin{equation}
	\hat{V}\ \ |\Gamma, a_ib_ic_i\rangle  = {1\over 4}\ l_{P}^3\
\sum_i
	\sqrt{\left |a_ib_ic+a_ib_i+b_ic_i+c_ia_i
	\right | }\ \  |\Gamma, a_i b_i
	c_i\rangle,
\end{equation}
where we restored the suffix $i$ that indicates the node to which
$a, b$ and $c$ refer. We recall
that the sum is over all nodes $i$ in the region
$R$.
In other words, the component of the spectrum of the volume of the
region $R$ corresponding to the trivalent states is the one given in
(1).

Finally, let us consider the non-trivalent states.  The nodes of order
larger than 3
are not unique, thus in general the coloring $p_l$ of the link are not
sufficient to characterize the state and we need an additional set of
integers $v_i$, one for every node $i$, to characterize the
routing of the individual segments through the $i$-th node.
 We may thus indicate the states as $|\Gamma, p_l,
v_i\rangle$.
The
range of the index $v_i$ is finite, and depends on the order of the
$i$-th
node and on the coloring of the adjacent links. For instance,
there
are two independent 4-valent nodes with all 4 links colored 1; thus
for such an node $v_i=1,2$.

If the point $P_I$, to which the $I$-th  box is shrinking in the limit,
coincides with
the $i$-th node, and this node is not trivalent, then
it is
not true anymore that in the limit the grasping gives the original
state
back, precisely because nodes of order larger than 3 are not
unique.  A rearranging of the routing of the segments at the
node
may modify the state.  Thus, equation (\ref{trivalent}) does not hold
anymore.
However, since the possible routings $v_i$ through each node
are
of finite number, by repeating the same reasoning as in the trivalent
case, we have in general that
\begin{equation}
	\hat{V} \ |\Gamma, p_l, v_i\rangle  = \sum_i
	\sqrt{2^{\scriptscriptstyle -7} | \hat C_i | }
	 \ |\Gamma, p_l, v_i\rangle
	\label{operator}
\end{equation}
where the operator $\hat C_i $
\begin{equation}
	\hat C_i \  |\Gamma, p_l, v_i\rangle =
	\sum_{v_i'} C^{v_i'}_{v_i}\ |\Gamma, p_l, v_i'\rangle,
\end{equation}
is a finite dimensional matrix,  uniquely determined by the
order
of the $i$-th node and the coloring of its adjacent links.
These matrices are defined by
 \begin{equation}
	\sum_{v_i'} C^{v_i'}_{v_i}\ |\Gamma, p_l, v_i'\rangle=
	\sum_m  \epsilon_m
	\sum_{l_\sigma l_\tau l_\rho}
	\sum_{STR}
 	\sum_{j=1,8}
	|(\alpha_{ \sigma\tau\rho}
	\#_{(s_St_Tr_R)} \gamma^{(\Gamma, p_l)}_m)^j\rangle
\end{equation}
and can be computed using the Penrose calculus, as we did above.

Clearly, the eigenstates of the volume operator can be found by
diagonalising the $C^{v_i'}_{v_i}$ matrices.  If we denote by
$c^{(n)}_{v_i}$ the eigenvectors and by $\lambda_i^{(n)}$ the
eigenvalues, so that $C^{v_i'}_{v_i}\ c^{(n)}_{v_i'} = \lambda_i^{(n)} \
c^{(n)}_{v_i}$,
then it is immediate to verify that the states
\begin{equation}
	|\Gamma, p_l, n_i\rangle = \sum_{v_1...v_I}\ c^{(n_1)}_{v_1} ...
c^{(n_I)}_{v_I}\ |\Gamma, p_l, v_i\rangle
\end{equation}
where $I$ is the number of nodes in $\Gamma$, are eigenstates of
the volume with eigenvalue
\begin{equation}
 V = \sum_i \sqrt{2^{{\scriptscriptstyle{-7}}} |\lambda_i^{(n_i)} | }.
\end{equation}
This formula gives the complete spectrum of the volume operator. To
compute it explicitly, we must calculate and diagonalise the family
of matrices $C^{v_i'}_{v_i}$.

We conclude this section with a most important observation.
Consider the case in which the region $R$ is given by the entire
three dimensional space.  In this case, it is immediate to verify
that the quantum volume operator, defined in (\ref{operator}),
satisfies
\begin{equation}
	 U(\phi) \hat V = \hat V U(\phi)
\end{equation}
for any diffeomorphism $\phi$ of the three-manifold (in the
connected component of the identity), where $U(\phi)$ is the
generator of finite diffeomorphisms which acts on the quantum
state by displacing loops \cite{lp1}. This is an immediate
consequence of the fact that the action of $\hat V$ is only sensitive
to features of the loop state that are invariant under
diffeomorphisms.  It follows that $\hat V$ is well defined on the
equivalence classes of quantum states under diffeomorphisms,
namely on the knot states \cite{spinnet}.   Thus $\hat V$ defines a
genuine operator acting on knots states.  This operator acts on the
intersections of the knots, with no reference to position (or
momentum)
space. It represents an example of the combinatorial operators
which may describe quantum general covariant physics on knot
space.

\section{Area}

The operator $\hat{A}[{\cal S}]$ that corresponds to the area of a
surface ${\cal S}$ was defined in \cite{lp4},
where a component of its spectrum was computed.  This component
corresponds to the eigenstates formed by states in which the
surface ${\cal S}$ is pierced by single lines, or 1-ropes.  The rest of
the spectrum, which we are about to compute, corresponds to
states in which the surface is crossed by ropes of degree higher than
one.

The area of a surface ${\cal S}$ is given, in terms of the Ashtekar
conjugate variable by \cite{area}
\begin{equation}
	A = \int_{\cal S} d^2\sigma\  \sqrt{\tilde E^{ai}\tilde
	E^{bi}\, n_a n_b}
\end{equation}
The two-hands loop observable
\begin{equation}
	T^{ab}[\alpha](s,t) = Tr[U_\alpha(s,t)\,\tilde
	E^{a}(\alpha(t))\,U_\alpha(t,s)\,\tilde E^{b}(\alpha(s))]
\end{equation}
converges to $16 \tilde E^{ai}(x)\tilde E^{bi}(x)$ when $\alpha$
shrinks to $x$. Therefore for a small surface ${\cal S}_I$ we can
write
\begin{equation}
	A_I^2 = \int_{\cal S_I} d^2 \sigma
	\int_{\cal S_I} d^2 \tau
	\left| {1\over 8}\ n_a( \sigma)n_b(\tau)
	T^{ab}[\alpha_{ \sigma \tau}](s,t)\right|
	\label{a2}
\end{equation}
up to small terms; here $\alpha_{\sigma\tau}$ is a small loop
going through $\sigma$ and $\tau$.  Following the same idea used
for the volume, we
partition the surface $\cal S$ into small square surfaces  ${\cal
S}_I$ of side $L$, so that we have
\begin{equation}
	A = \lim_{L\rightarrow 0} \sum_I \sqrt{ A_I^2 },
\end{equation}
and we define the quantum area operator as
\begin{equation}
	\hat A = \lim_{L\rightarrow 0} \sum_I \sqrt{\hat A_I^2 },
\end{equation}
where $\hat A_I^2$ is defined by replacing $T^{ab}$ with  $\hat
T^{ab}$ in (\ref{a2}).

Let us now compute the action of $\hat A$ on a spin network state.
{}From the definition of the two-hands loop operator \cite{lp1}
we have
\begin{eqnarray}
	\hat A_I^2  \ |\Gamma, p_l\rangle =&&
	\int_{\cal S_I\times\cal S_I}
	d^2 \sigma\, d^2 \tau
	\left| {1\over 8}\ n_a( \sigma)n_b( \tau)
	\sum_m \epsilon_m
	\int ds \int dt
	\right.
\nonumber \\
	&& \times
	\left.
	\dot\gamma_m^a(s)\,\delta^3(\gamma_m(s), \sigma)\
	\dot\gamma_m^b(t)\,\delta^3(\gamma_m(t), \tau)\
	n_a( \sigma)n_b( \tau)
	\right|
\nonumber \\
	&& \times
	\sum_{j=1,4}
	|(\alpha_{ \sigma \tau}\#_{st}\gamma_m)^j\rangle.
 \end{eqnarray}
The right hand side gets contributions only from the intersections
between $\gamma_m$ and the surface.  For small enough $L$ there
will be at most one rope $l$ crossing ${\cal S}_I$. Assume that this
is the case and that the rope $l$ has degree $p$, and label its
segments by an index $P$.  Following the same argument we used for
the volume operator, we have
\begin{eqnarray}
	\hat A_I^2  \ |\Gamma, p_l\rangle =&&
	\int_{\cal S_I\times\cal S_I}
	d^2 \sigma\, d^2 \tau
\nonumber \\
	&& \times
	\left| {1\over 8}\ n_a( \sigma)n_b( \tau)
	\sum_m \epsilon_m
	\int ds \int dt \right.
\nonumber \\
	&& \times \left.
	\dot l^a(s)\,\delta^3(l(s), \sigma)\
	\dot l^b(t)\,\delta^3(l(t), \tau)
	\right|  \ \ |\Psi\rangle
 \end{eqnarray}
where
\begin{equation}
	|\Psi\rangle =  \sum_{P}\sum_{P'}
	\sum_{j=1,4}
	|(\alpha_{ \sigma \tau}
	\#_{s_Pt_{P'}}\gamma_m)^j\rangle
	= c(p)\ |\Gamma, p_l\rangle.
\end{equation}
 (If the intersection point between the surface and the spin network
is a node, then only the loops rooted through the node accross the
surface contribute to the sum.) Inserting in the definition of the
area operator we have
\begin{eqnarray}
	\hat A \ |\Gamma, p_l\rangle &=&
	\lim_{L\rightarrow 0} \sum_I
	\left(
	\int_{\cal S_I\times\cal S_I}
	d^2 \sigma\, d^2 \tau
	\left| {1\over 8}\ n_a( \sigma)n_b( \tau)
	\int ds \int dt  \right. \right.
\nonumber \\
	&& \times \left.\left.
	\dot l^a(s)\,\delta^3(l(s), \sigma)\
	\dot l^b(t)\,\delta^3(l(t), \tau)\
	c(p)
	\right|\ \right)^{-1/2}  \  |\Psi\rangle.
 \end{eqnarray}
The integrals are immediate, giving
\begin{equation}
	\hat A \ |\Gamma, p_l\rangle =
	\lim_{L\rightarrow 0} \sum_I
	\sqrt{\left| {1\over 8}\ c(p) \right|} \ \ |\Psi\rangle.
 \end{equation}
Two kinds of terms come into the evaluation of this action.
The combinatorial calculus is summarized in Figures 12 and 13.
\begin{figure}[t]
\epsffile{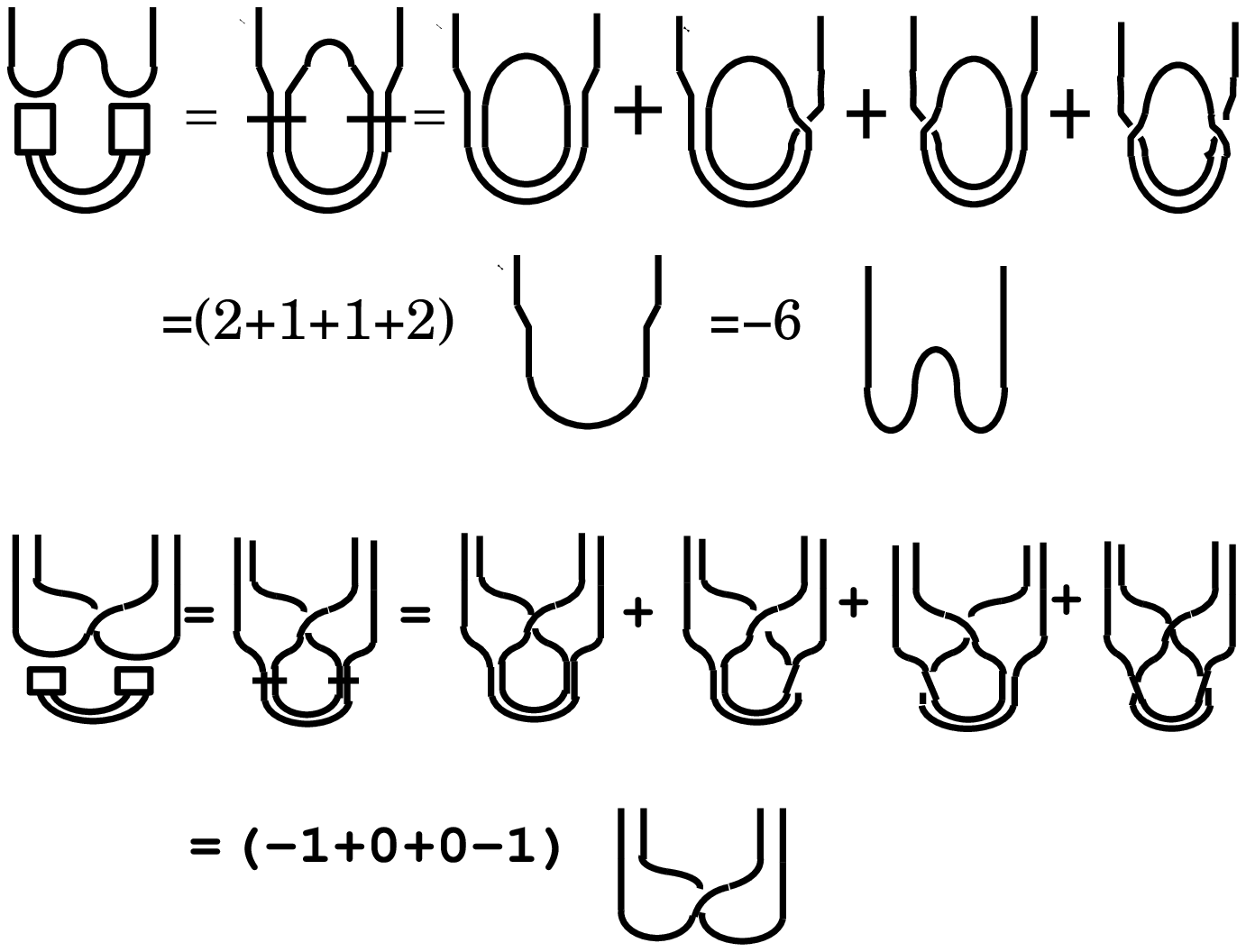}
\caption{The two kinds of terms that come into the action of the area
operator}
\end{figure}
\begin{figure}[t]
\epsffile{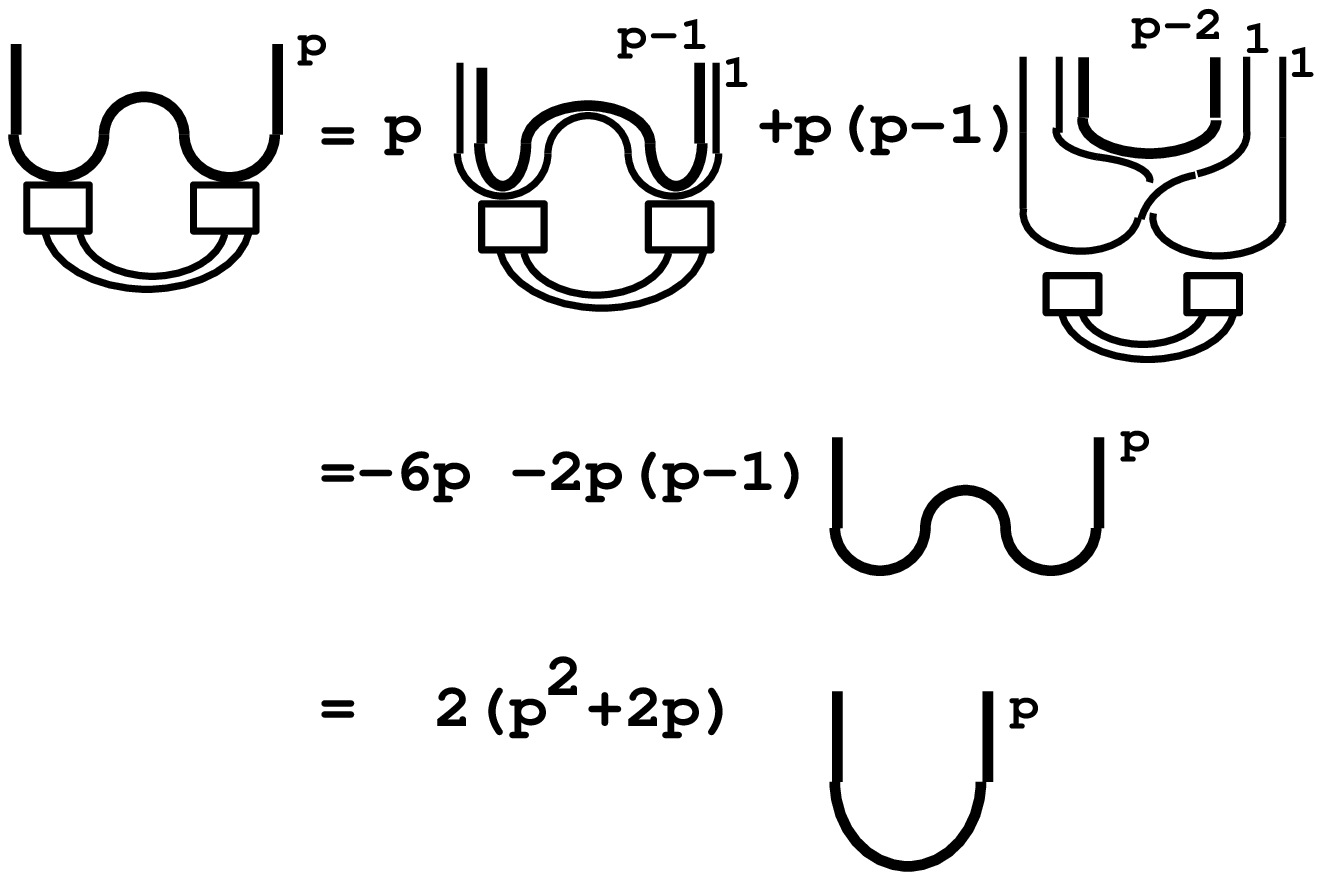}
\caption{The calculation of the action of the area
operator}
\end{figure}
The result is
$c(p)=-2(p^2+2p)$, so that we can take the limit and conclude
\begin{equation}
	\hat A |\Gamma, p_l\rangle =  {1\over 2} \ l_P^2
	\sum_l  \sqrt{p^2+2p} \ |\Gamma, p_l\rangle
 \end{equation}
where the sum ranges over all links $l$ crossing the surface. The
spectrum given in eq.(2) follows immediately.\footnote{This
formula corrects the numerical factor reported in \cite{lp4},
which was incorrect due to a miscounting of trace factors.}

This result admits a surprisingly simple interpretation.  Recall
that a rope $l$ of color $p_l$ corresponds to a parallel propagator
(in the connection representation) in the $j_l=p_l/2$
representation, and take units in which the Newton constant $G$
is one, so that $\l_P^2 = \hbar$. Then we can write the area of the
surface as
\begin{equation}
	{\cal A} =  \sum_l  \hbar\ \sqrt{j_l(j_l +1)}.
\end{equation}
Since the $\hbar\,\sqrt{j(j +1)}$ is the angular momentum $L$
associated to the $j$-th representation of $SU(2)$, we have that,
\begin{equation}
	{\cal A} =  \sum_l  L_n  =  L_{\rm total}
\end{equation}
where $L_n$ is the angular momentum of the $SU(2)$
representations associated with each rope crossing the surface.
See reference \cite{area}.  This remarkable fact has been pointed
out by J. Iwasaki \cite{junichi}.

\section{Spectra as physical predictions}

Can we relate the spectra that we have computed to physical
measurements of real volumes and areas?  If we want to measure
the volume of a region $R$, we have to physically identify, or
specify, the region $R$ in some way or another.
If we disregard the relativistic properties of the gravitational field,
a region $R$ can be identified by means of physical objects, and we
can assume that these objects are dynamically decoupled from the
gravitational field.  As is well known, this
procedure cannot be extended to the relativistic gravitational
theory: there is no physical object whose dynamics is decoupled
from gravity, nor, actually, is there any meaning to such a
decoupling.  In a relativistic gravitational context we must regard
any object used to identify a physical region as dynamically coupled
to the gravitational field \cite{whatiso}.  Therefore the volume and
area that we measure are always determined by objects
(in a field theoretical perspective, fields) which are components of
the dynamical theory.

The remark, which we believe represents the heart of the
physical meaning of general
covariance\footnote{or, equivalently, diffeomorphism invariance.},
is reflected in the
mathematics of the theory as follows. The volume and area
observables $V$ and $\cal A$ considered in the previous sections are
defined in pure gravity and thus are functionals of the sole
gravitational degrees of freedom: $V[g], {\cal A}[g]$.  A such, they
do not commute with the canonical constraints, that is, they do not
transform as singlets under 4-dimensional diffeomorphisms of the
dynamical variables on which they depend ($g$); therefore they are
not physical observables of the theory.  However, if matter fields
$\phi$ are coupled to the theory, and if we
consider volumes $V_{Ph}$ and areas ${\cal A}_{Ph}$ of regions {\it
specified by those matter
fields,} then such volumes and areas do commute with the
constraints. Indeed $V_{Ph}=V_{Ph}[g, \phi]$ and ${\cal
A}_{Ph}={\cal A}_{Ph}[g, \phi]$ are functionals of
the gravitational field as well as of the matter fields, which
determine the domain of integration; a diffeomorphism of all the
dynamical variables generated by the constraints acts then on the
integrand as well as the integration domain.  See
\cite{whatiso,mc,mc2,matterrefl} for examples.
The observables  $V_{Ph}$ and ${\cal A}_{Ph}$ that express
volumes and areas of regions
physically identified by dynamical matter are the  quantities in the
theory that truly correspond to the volumes and areas that we
concretely measure.  Our problem is therefore to understand the
relation between the spectra of $V$ and ${\cal A}$ computed above
and the spectra of  $V_{Ph}$ and ${\cal A}_{Ph}$.   We are now going
to argue that the second ones coincide with the first ones.

We may begin this discussion by noting that
we can always  fix (perhaps partially) the
diffeomorphism gauge invariance by constraining
the values of certain matter fields
at given spacetime coordinates or, equivalently,
by defining preferred spacetime coordinates in terms of these matter
fields.  In the gauge fixed theory, the matter degrees of freedom
that are employed in the gauge fixing
disappear, but the gauge freedom is also restricted, so that the total
number of degrees of
freedom of the theory remains the same.
Let us sketch two examples.

First, consider a partial gauge fixing. In order for the
volume of the universe (assuming this is spatially compact) to have
any meaning at all in the general
case in which there are no symmetries, we have to specify
a synchronization of clocks; so that a spacelike surface, having a
volume, is
singled out.  However we achieve this concretely, we can
formally represent the
field of clocks by means of a scalar field $\phi(\vec x, t)$, whose
dynamics reproduces the evolution of the clocks in coordinate time.
Let then $R_T$ be the region
of the points where the clocks indicate $T$, namely where the scalar
field has value $T$.  The volume of $R_T$ is given by
\begin{eqnarray}
	V_{Ph} &=& \int_{R_T}
	\sqrt{\det {}^3g}
\nonumber \\
	&=& \int d^4x\ \delta^4(\phi(x), T)\ \sqrt{\det {}^4g\
	{}^4g^{\mu\nu}\partial_\mu\phi\partial_\nu\phi}
\nonumber \\
	&=& V_{Ph}[g,\phi],
\end{eqnarray}
where we have indicated by ${}^3g$ the spacial metric and by ${}^4g$
the 4-dimensional one.
It is easy to verify that $V_{Ph}$ commutes with all the constraints
(including the hamiltonian constraint) \cite{whatiso}; indeed, this is
a ``physical volume", in
the sense defined above.   Now, the point is that we can gauge fix the
hamiltonian constraint by requiring that $\phi(\vec x, t) = t$. In this
gauge we obtain a highly non-trivial hamiltonian, which depends on
the actual dynamics of $\phi$, and the only remaining constraints
are the diffeomorphism constraint.  Thus, in this gauge the theory
is formally equivalent to pure GR, but with no hamiltonian
constraint and a non-trivial hamiltonian. With respect to the
preferred physical time coordinate that we have selected, the
3-dimensional geometry of space becomes a
physical observable; notice that the 3-dimensional geometry
contains 3 degrees of freedom, which are precisely the 2 degrees of
freedom of the pure gravitational field plus the 1 of the scalar field,
which has disappeared.  Now, in the gauge fixed theory the
gauge-invariant observable quantity $V_{Ph}$ is given by
\begin{equation}
	V_{Ph} = \int d^3x\  \sqrt{\det {}^3g},
\end{equation}
But this observable has precisely the form of the
non-gauge-invariant volume (\ref{int}).  This gravity+matter gauge
fixed theory admits a quantization fully parallel to the pure gravity
theory, only the hamiltonian constraint has now been replaced
by a hamiltonian.
However, nothing in the construction of sections 2 and 3 depend on
how the dynamics is to be treated.  Therefore the quantum operator
representing $V_{Ph}$ in the gauge-fixed theory with matter has the
very same form as the one representing $V$ in the pure gravity
theory. Therefore the two have the same spectrum.

As a second example, consider the case in which we model a full
material reference system by means of four scalar matter fields
$\phi^\mu(x)$, with some fixed dynamics, and we gauge fix the
theory entirely by defining coordinates $x^\mu =\phi^\mu(x)$.  The
case in which such a matter describes dust has been recently
analyzed in great detail by Brown and Kucha\v{r}
\cite{karel} from the point of view of
canonical quantization.  In the fully gauge fixed theory there
is
no remaining constraints (there is a highly non-trivial hamiltonian).
The full gravitational field, as a function of the preferred
coordinates, is observable.  If we now fix a spacial region
$R$ in the (preferred) coordinate space, the volume of any region $R$
is observable.  It represents the volume of a region identified by
material objects described by the $\phi^\mu(x)$ fields.  Again, in the
gauge fixed theory the operator that corresponds to the volume of
$R$ is precisely the volume operator studied in the paper.

At first sight, the result of the above discussion has a
counter-intuitive aspect. Indeed, one may be tempted to object that
while quantum properties of gravity can yield the quantization
of, say, the volume of a region $R$, we must take into account
the possibility that the
matter identifying $R$ must also be quantized  and thus subject to
quantum fluctuations.  Thus, the discrete spectra found here might
be smeared out into a continuum by the quantum fluctuations of
the matter fields by which the region $R$ is defined.

However, a more careful analysis shows that this
consideration is wrong.  The reason is that it neglects
diffeomorphism invariance. Indeed, the mistake is to
assume that the matter that identifies the region {\it as well as\ }
all of the components of the metric tensor, which determines the
volume of the region, independently undergo quantum
fluctuations.  This is wrong because all these component fields,
being not gauge invariant, cannot represent independent
degrees of freedom of the quantum theory.  Instead, only the gauge
invariant quantities that represent true
dynamical degrees of freedom are subject to the quantum
uncertainty principle and to quantum fluctuations.
Therefore, in a general covariant theory, the kinematical volume by
itself is not a gauge invariant degree of freedom, nor is the value of
the matter fields in a coordinate region.  The dynamical variable
that expresses the volume is a combination of the two, and it is only
this combination which is dynamically meaningful and which is
quantized.  One can pick a gauge by fixing the coordinates to the
values of the matter fields, in which case it appears that
the dynamical quantum degrees of freedom are represented
by certain components of the metric.   Or one could
fix components of the metric, so that the matter fields represent,
in such a gauge, the dynamical degrees of freedom.  But
these are
equivalent descriptions of the same physics.  What we cannot
do is pick a gauge which allows us to view
both the metric and the reference matter as independent quantum
objects.   To put it differently, reference
matter is not quantized, in a gauge in which it is used up by the
gauge fixing,  for the
same reason for which the Higgs field is not treated as
a quantum field in those gauges in which its degree of freedom
is represented by the longitudinal components of the
gauge bosons.

In conclusion, since the calculations of sections 2 and 3 do not
depend on the dynamics determined by the hamiltonian constraint,
we can reinterpret these calculations as referring to physical areas
and volumes determined by reference matter.  Therefore, as we have
emphasized in the introduction, we may conclude
that the the spectra
of volume and area computed here may
be considered to constitute physical predictions
of the quantum theory of gravity in the loop representation.

\section{Conclusions}

Many different approaches to quantum gravity incorporate a
fundamental length.   Well-known semiclassical arguments, some
simple, other more sophisticated, suggest that geometrical
measurements must be limited by an intrinsic quantum
uncertainty of the order of the Planck length, in quantum general
relativity \cite{short}, as well as in string theory \cite{string}. In
string theory, in particular, the existence of a short scale discrete
structure has been indicated as a possible reason for the exponential
damping of high energy scattering amplitudes \cite{he} and for the
slower growth of the thermal partition function at high temperature
\cite{witten}, and the intriguing short scale string discreteness
discovered by Klebanov and Susskind \cite{ks} has been recently
indicated by Horowitz as a suggestive indication of closeness
between string theory and loop representation \cite{gary}. (On the
relation between string theory and loop representation, see
\cite{jb}.)  A relation between black hole entropy and area
quantization was suggested by Bekenstein \cite{bek} and recently
studied by Jacobson \cite{jac}, and some kind of short scale
discreteness is implicit in the idea of spacetime foam \cite{foam}.
For an interesting and
comprehensive overview of many of these ideas on a fundamental
length, see the recent review \cite{garay}, which concludes noticing
that ``the presence of a lower bound to the uncertainty of geometry
measurements seems to be a model-independent feature of quantum
gravity".

Here, such a lower bound, and its associated spectral discreteness,
derive from the straightforward non-perturbative quantization of
general relativity.  The discreteness is specific, as we have
detailed, non-trivial formulae for the spectra of volume and area
(equations (1) and (2)).

A characteristic aspect of our results is that they are entirely
kinematical.  As the Hamiltonian constraint plays no role in the
derivation, the results must be independent of the dynamics of general
relativity.  In this respect, what we have done here may be
considered to have carried quantum gravity to the same point that
normal ordering carried quantum electrodynamics.  In that case, the
correct treatment of a kinematical operator product divergence
results in the discovery of the physical content of the theory:
photons, electrons and positrons.  The stage is set for the definition
of dynamics via the hamiltonian.  Here, a more complicated
procedure, made necessary by the requirement of
diffeomorphism invariance and the absense of a background metric
and an associated positive/negative frequency distinction, reveals the
physical content of the theory: the discrete states labeled by the
spin networks. Again, what remains is to apply the dynamics to these
states, either via the Hamiltonian constrant or an appropriately
defined hamiltonian\cite{lp5}.

The fact that the results found here derive only
from a succesful treatement of kinematical, operator product,
divergences, gives them a certain
robustness.  To the extent that the results here
stand as predictions of quantum general relativity, they are also
predictions of quantum supergravity, or any of the large number
of modifications of general relativity such as those that involve
dilaton fields or higher derivative terms.

The  discreteness of areas and volumes found here then supports
and strengthens the
evidence for the existence of a discrete short-scale structure of
space, which  emerged first from studies of the kinematical
 ``weave'' states\cite{lp4,jun1}.  However, these
results go further in
extending these results to the diffeomorphism invariant
level, and in providing a precise physical meaning to
the claim of the existence of a short-scale structure, one tied to
the topological discreteness of knot space.  This might have
far reaching consequences for our understanding of the general
structure of infinite-dimensional, diffeomorphism invariant
quantum field theories.

It is important to note that the framework here is then
substantially dissimilar from the conventional framework in which
discreteness of the regularized theory is scaled away by the large
scale limit taken at a critical point that defines a continuum theory.
Here the
discreteness is physical. As a result, the
relation between short distance and long distance regimes of the
theory is very different than in quantum field theories defined on
metric manifolds.  The continuum limit must be a limit in which
universes that are large on the Planck scale, and behave
semiclassically, are constructed, not in which a short distance
cutoff is taken to zero.

This does not mean that further divergences
might yet be encountered in the evaluation of the dynamics
of the theory.     However, as the
handling of the kinematical divergences has, in this case, revealed a
state space with a natural discrete structure and short distance
cutoff, we do not expect anymore the appearance of the conventional
ultraviolet divergences, which come from summing over the free
Fock states of arbitrarily short wavelength.  Here the possible
sources of divergences  correspond to
infinitely large networks, or to arbitrarily large values
of the labels on the graphs.
But, by the results found here, these correspond to limits of
large volumes and large areas; hence these must be infrared
divergences.

There remains still much to be understood about these issues.  In
particular, it is important to note that the Planck
constant $l_P$ appearing in the spectra (1) and (2) is a bare
quantity, that may very well suffer finite renormalizations and thus
not coincide with the macroscopic value of
$\sqrt{\hbar G_{NEwton}/c^3}$.\footnote{For an
indication that this does occur, see \cite{lp5}.}

Finally, we must mention that we can make
no claim that the spectra computed here are
unavoidable consequences of just quantum theory and general
relativity, since choices and assumptions are made in the
construction of any specific quantum theory.  Assumptions about the
nature of the state space on which the theory is constructed may
affect the theory substantially.  Ashtekar and others
\cite{math,math3} are developing a deep mathematical analysis of
the loop representation capable of addressing these issues. A case to
be examined in this regard is the extended loop representation
\cite{extended}: it will be interesting to see if the present results
can be derived in that case as well.   On the other hand, the spectra
computed here make the present form of the loop representation
theory falsifiable, in spite of its incompleteness.

In the absence of Planck scale measurements, it is of course
hard to imagine how predictions for the discreteness of these
observables could be tested.  However, the examples of the
early developments of
quantum mechanics and solid state physics suggests that this
discreteness could have implications for the thermodynamics of the
gravitational field\cite{termo}.  This could have important
implications for the interpretation of black hole entropy, as well
as for the question of the production, in the very early universe,
of a spectrum of primordial gravitational radiation.  These
questions are presently under investigation.

\vskip 2cm

We are grateful to Roger Penrose for explaining spin networks to us
and for advise about these calculations, and to Roumen Borrisov for
checking our calculations in detail and spotting a wrong factor of 2.
We thank Jan Ambjorn, Abhay Ashtekar, John Baez, Bernie
Bruegmann, Mauro Carfora, Louis Crane,
Josh Goldberg, Gary Horowitz, Chris Isham, Junichi
Iwasaki, Ted Jacobson, Karel Kucha\v{r}, Renate Loll, Alexander
Migdal, Jorge Pullin and Ranjiv Tate, for comments, suggestions,
conversations and criticisms.   Finally, we thank the Isaac Newton
Institute in Cambridge where this work was conceived, and
SISSA, in Trieste, where some of it was carried out.  This work
was partially supported by the NSF under grants PHY539634,
PHY9016733 and INT8815209 and by research funds of Penn State
University.

\end{document}